\documentclass[preprint]{aastex}
\usepackage{emulateapj5}
\usepackage{epsfig}

\begin{document}

\title{On the formation of an eccentric disk via disruption 
of a bulge core near a massive black hole}

\author{A.~C. Quillen \& Alex Hubbard
\affil{Department of Physics and Astronomy, 
University of Rochester, Rochester, NY 14627; 
{\it aquillen@pas.rochester.edu, hubbard@pas.rochester.edu}
 }
}

\begin{abstract}
We consider the possibility that an infalling bulge core  or
stellar cluster could form an eccentric disk, 
following tidal disruption by a massive black hole
in the center of a galaxy.
As a function of central black hole mass,
we constrain the core radii 
and central densities of cluster progenitors capable of
becoming nearly Keplerian disks which can 
support lopsided slow modes.
We find that progenitor stellar clusters with core radii less than a pc
and densities above a few times $10^5 M_\odot {\rm pc}^{-3}$
are likely eccentric disk progenitors 
near a massive black hole of mass $10^7$ -- $10^8 M_\odot$.
Lower density and larger progenitor cores are capable of causing 
eccentric stellar disks near more massive black holes.
Our constraints on the progenitor cores are consistent
with existing N-body simulations, which in one case
has produced an eccentric disk.   
For M31 and NGC 4486B, the estimated progenitor cluster
cores are dense and compact
compared to Galactic globular clusters,
however the cores of nearby galaxies such
as M33, M32 and M31 itself are in the right regime.
If galaxy mergers can create
eccentric disks, then they would be a natural
consequence of hierarchical galaxy formation.

\end{abstract}

\section{Introduction}

Ever since the double-peaked nucleus of M31 was clearly
resolved by Hubble Space Telescope (HST) imaging \citep{lauer}, 
the morphology of this system
has been a challenge to explain.  The most successful
kinematic model is the eccentric stellar disk proposed by
\citet{tremaine} where stars in apsidally aligned elliptical orbits 
about a massive black hole result
in an over-density at their apoapses, thus accounting for 
the brighter peak P1.  The black hole itself
resides near the fainter, more centrally
located peak denoted P2.
Though this model has been successful at matching
the observed velocity and luminosity distribution
(\citealt{kormendy,statler,bacon,sambhus}),
eccentric disk formation remains a mystery.

\cite{touma} showed that a small fraction of counter rotating stars,
possibly originating from an disrupted globular cluster
on a retrograde orbit, 
could cause a more massive pre-existing stellar 
disk to develop a lopsided or $m=1$ instability.
The self-consistent kinematic modeling of 
\citet{sambhus}, which requires a small percentage
of counter rotating stars, supports this scenario.
\cite{bacon} proposed that a collision from a passing
molecular cloud or globular cluster could knock a pre-existing
stellar disk off-center, resulting in a long lived,
precessing mode.  The N-body simulations of \citet{jacobs,taga}
have illustrated that lopsided stellar disks could be long-lived.

The scenarios discussed by \citet{sambhus,touma,jacobs,taga,bacon},
begin with an initially axisymmetric stellar disk, which then
becomes lopsided either because of a violent event 
(such as a collision with a globular cluster) or
the growth of an instability.  However \citet{bekki} 
proposed that M31's eccentric disk could
have been a result of a single disruption event.
He succeeded at producing an eccentric stellar
disk by disrupting a globular cluster near a massive black hole
in an N-body simulation, though an SPH simulation of the
disruption of a massive gaseous cloud did not yield
an eccentric disk \citep{bekki2}.
The massive black hole in in M31 is moderate
with a mass $3 -7 \times 10^7 M_\odot$ \citep{kormendy,bacon}.
The eccentric disk itself is nearly as massive as the black hole,
$\sim 3 \times 10^7 M_\odot$
\citep{peng, sambhus, bacon}, presenting
a problem for the globular cluster disruption scenario
proposed by \citet{bekki} which assumed that the globular
cluster was typical of Galactic globular clusters
and of order a million solar masses.


Here we reconsider Bekki's proposal, that a single
disruption event could have produced the eccentric disk in M31.
The disruption event is likely to be complex so
we focus on what final disks are capable of supporting
lopsided a slow modes, drawing from 
the recent work of \citet{tremaine2001} who developed a formalism 
for the purpose of predicting the precession
rates and eccentricities of 
discrete modes for low mass, nearly Keplerian disks.

We place limits on the density and core radius
of progenitor clusters for eccentric disks near massive black holes.
The resulting diagram is useful
toward predicting what types of galaxies are most likely
to harbor eccentric disks, for predicting
the probability that eccentric disk formation events occur and
for guiding the initial conditions of N-body simulations which may 
determine if and how they form.

\section{ Tidal disruption of a cluster }

The observed correlation between mass of a black hole and the bulge dispersion 
\citep{gebhardt2000,ferrarese} allows us to relate the black
hole to the bulge of its host galaxy.
\begin{equation}
\log(M_{bh}/M_\odot) = (8.13 \pm 0.06) + (4.02\pm 0.32)\log
\left({
  \sigma_*\over 200 {\rm km~s^{-1}}
  }\right)
\end{equation}
where $\sigma_*$ is the stellar bulge dispersion, and 
we adopt constants in the relation given by \citet{tremaine2002}.
For a discussion on the systematics of measuring
the velocity dispersion within $r_e$, the effective or
half light radius, see \citet{tremaine2002}.


The transition radius (sometimes called the sphere of influence) 
where the gravity
from the black hole takes over from the bulge is
\begin{equation}
\label{rt_def}
r_t =   {GM_{bh} \over 2 \sigma_*^2} =  0.5 {\rm pc }
          \left({M_{bh} \over 10^7 M_\odot }\right)
          \left({200 {\rm km/s}  \over \sigma_*  }\right)^2
\end{equation}
Using the above relation between the bulge and the black hole mass,
\begin{equation}
\label{r_t}
r_t \approx 1.8 {\rm pc}
          \left({M_{bh} \over 10^7 M_\odot }\right)^{0.5}.
\end{equation}

We now consider disruption of a stellar cluster or molecular
cloud by a massive black hole.
The tidal force from the black hole and bulge is 
\begin{equation}
F_{tidal} = 
\left[{2 G M_{bh} \over r^3} + {2 \sigma_*^2 \over r^2}\right] s
\end{equation}
at distance $r$ from the galaxy nucleus and distance
$s$ from the cluster nucleus.
We set this equal to the gravitational force from the
cluster on itself to determine how the cluster disrupts. 

We describe the nucleus of a cluster with two parameters,
the core radius, $r_0$,  and the density within this
radius, $\rho_0$.  The cores of isothermal spheres and King models
are characterized with with these two parameters (e.g., \citealt{B+T}).
Within the core radius
of the cluster, we assume that the cluster has 
nearly constant density, so
its gravitational potential is proportional to $s$.
We set the tidal force from the galaxy equal to that from the cluster and
determine the distance $r_d$, (the disruption radius) 
when the two are equal.
The distance from the galaxy nucleus where the entire core
of the cluster disrupts is then set only by the cluster core
density, $\rho_0$.
\begin{equation}
{G M_{bh} \over r^3 } + {\sigma_*^2 \over r^2} \sim G\rho_0
\end{equation}
Inside the transition radius the disruption radius
\begin{equation}
\label{inside}
r_d \sim  \left({M_{bh}\over \rho_0}\right)^{1/3} =  
          4.6 {\rm pc} 
          \left({M_{bh} \over 10^7 M_\odot}\right)^{1/3}
          \left({ 10^5 M_\odot \rm{pc}^{-3} \over \rho_0}\right)^{1/3}.
\end{equation}
Outside the transition radius
\begin{equation}
\label{outside}
r_d \sim  
\left({\sigma_*^2\over G\rho_0}\right)^{1/2} = 
          4.7 {\rm pc} 
          \left({\sigma_* \over 100{\rm km/s}}\right)
          \left({ 10^5 M_\odot \rm{pc}^{-3} \over \rho_0}\right)^{1/2}.
\end{equation}

We expect that it is difficult to make a slowly precessing eccentric
disk in a region where the black hole does not dominate the gravitational
field.  Outside the sphere of influence, 
the potential is strongly non-Keplerian.
The requirement that the cluster be disrupted (survive to) within 
the transition radius is  $r_d < r_t$.
Using the previous equation,  the definition for $r_t$,
and the relationship between the bulge
stellar dispersion and the black hole mass this constraint
gives the requirement 
\begin{equation}
\label{rholim}
\rho_0 \gtrsim 7.4 \times 10^5 M_\odot {\rm pc}^{-3} 
          \left({ M_{bh} \over 10^7 M_\odot }\right)^{-0.5}.
\end{equation}
From this we see that the
less massive the black hole, the more dense a cluster would be required
to form an eccentric disk.
This condition is our first constraint on progenitor clusters 
for eccentric disks.

At radii larger than the disruption radius, the outer parts
of the stellar cluster or galaxy core would be stripped.  We expect that 
the radial distribution of the resulting disk will depend
upon the concentration of the cluster or radial profile
of the infalling galaxy bulge or core.  Since the 
densest part of the cluster is disrupted at $r_d$, we expect
that this part is most likely to be involved in any large
eccentricity amplitude variation.  So we can estimate the
mass of the active disk formed at disruption using 
the core radius of the cluster, $r_0$,  resulting in 
an active disk of mass equal to the mass
within the core of the cluster, $M_c \sim \rho_0 r_0^3$.
The width of this active disk would be of the order of the core radius
of the cluster.

\citet{tremaine2001,lee} assume that the mass of the disk
is less than that of the black hole.  If we consider the active
part of the disk, this is even true in M31 
(the center of mass is closest the black hole).  
It's difficult to imagine a stable situation 
when the lopsided part of the disk is as massive 
as the black hole
so we arrive at another constraint on our progenitor cluster,
\begin{equation}
M_c < M_{bh}.
\end{equation}

It is unlikely that a thin disk will result
following disruption unless the disk Mach number, ${\cal M}(r)>1$
\citep{tremaine2001},
\begin{equation}
\label{Mach}
{\cal M} (r_d) \equiv {\Omega(r_d) r_d \over \sigma_c} > 1
\end{equation}
where  $\Omega(r_d)$  is the angular rotation rate
at the disruption radius $r_d$, and $\sigma_c$ is
the central cluster dispersion.

Following disruption of the cluster, the energy spread of the particles
is set by the cluster core radius, $r_0$, so we expect
that the resulting width of the disk $ \sim r_0$.  
The angular momentum spread at disruption is also set by
the size of the cluster core,
${\Delta J \over J_0} \sim {r_0 \over r_d}$ where
$J_0$ is the angular momentum of the bulk motion 
of the cluster at disruption.
Consequently
following disruption we expect a disk radial velocity dispersion 
at least
\begin{equation}
\sigma_c \sim  \sqrt{G M_{bh} \over r_d} {r_0 \over r_d}.
\end{equation}
This scaling is similar to that outlined by \citet{johnston}
for the disruption of dwarf galaxies.
Alternately we can use the relation between central
dispersion, core density and radius
for a King model ($\sigma_c^2 = 4 \pi G \rho_o r_0^2/9$). 
Using either estimate for $\sigma_c$ and inserting
into equation(\ref{Mach}),
assume that we are within the transition radius
so that 
$\Omega = \sqrt{GM_{bh} /r^3}$ 
we estimate 
\begin{equation}
{\cal M}(r_d) \sim { M_{bh}^{2/3} \over \rho_0^{2/3} r_0^2 }  \gtrsim
 1
\end{equation}
which is equivalent to requiring $M_{bh} > M_d$.

In summary we can constrain properties of an incoming cluster
which subsequently forms an eccentric disk:
\begin{enumerate}
\item $r_d < r_t$.  The disruption radius should be less than the transition radius.
\item $r_0 < r_t$.  The cluster core radius should be less than the transition radius.
\item $M_c < M_{bh}$.  The cluster core mass should be less than 
the black hole mass.  This constraint is equivalent to requiring 
$r_0 < r_d$ or ${\cal M} > 1$.
\end{enumerate}

We plot these three constraints as lines in Figure 1 for three 
different black hole masses, and as a function of cluster
core radius, $r_0$,  and density, $\rho_0$.  
The allowed region of parameter
space for a disrupted cluster to form an eccentric disk
is the upper left hand corner, corresponding to core radii
above about a pc, and core densities above about 
$10^5 M_\odot {\rm pc}^{-3}$. 

Also shown in Figure 1 are measured core radii
and central densities (assuming a mass to light ratio
of $M/L_V=5$ in solar units) 
for the Milky Way globular clusters 
based on the quantities 
compiled by \citet{harris}.\footnote{We used the updated 
catalog available at http://physun.physics.mcmaster.ca/~harris/mwgc.dat}

We also include on Figure 1, the massive globular
cluster G1 (based on core properties estimated
by \citealt{meylan}) and the cores of M31, M32 and M33,
based on density profiles 
(or limits in the case of M32) estimated by \cite{lauer98}.
Limits on the density and core radius are placed on the
figure 1 for the two densest galaxies 
(other than M31, M32, and M33) studied by \citet{faber}.
Higher angular resolution observations would be needed to
find out if more distant galaxies can have cores with the high
stellar densities of M31, M32 and M33.

We see from Figure 1 that Galactic globular clusters are
capable of forming eccentric disks if they are extremely
dense and have compact cores, and they are more likely to do so in
galaxy nuclei containing larger black holes.
However, to provide a plausible progenitor for an eccentric disk, 
a cluster must not only have a dense and compact core but
be massive enough to account for the mass of
the entire disk. Galactic globular clusters are not
massive enough to account for the disk in M31.

\subsection{Bekki's simulations}

We examine the properties of the simulated disruptions
done by \citet{bekki,bekki2} to see if they are consistent
with our proposed progenitor cluster limits.
In an N-body simulation, \citet{bekki} succeeded in producing 
a stellar eccentric disk
by disrupting a globular cluster of mass 0.1 times
the black hole mass, for $M_{bh} = 10^7 M_\odot$.
The bulge he adopted was described by an NFW profile with a scale 
radius of $100$pc. 
For this adopted bulge profile we calculate
that the mass enclosed within 10pc was less than a million
$M_\odot$ so that the entire disruption took place
within what we have called the transition radius,
where the gravity
from the black hole takes over from the bulge (equation \ref{rt_def}).
His simulated cluster disrupted at a radius of $r_d \sim 3$pc  from the black 
hole, and since the cluster core was much smaller than
this,  $r_0 \sim 0.4$pc,
the stellar system was capable of forming an eccentric disk.
We conclude that the success of his simulation
in producing the eccentric disk is consistent with the crude
limits we have placed on progenitor cluster properties.

The SPH simulation by \cite{bekki2}, however, produced a circular disk, 
not an eccentric one.
The gas cloud in this simulation was of similar mass to
that of the stellar cluster, \citep{bekki} but had a significantly
larger scale length of 10pc.  The cluster disrupted
at $r_d \sim 10$pc which is approximately the same
as the core radius of the gas cloud.  We suspect that
a gas cloud with a smaller core radius would have been capable of 
producing an eccentric disk.
Again, we find that the failure of this simulation
in producing the eccentric disk is consistent with the crude
limits we have placed on progenitor cluster properties.

\subsection{M31's eccentric disk and possible progenitor
 cluster or galaxy core properties}

The two peaks observed in the nucleus of M31 
are separated by  1.87 pc \citep{lauer}
and the black hole mass is $3 \times 10^7 M_\odot$
\citep{kormendy}, so we expect a transition radius of $r_t \sim 3.1$
pc which is, as we expect, 
outside the location of the eccentric part of the disk.
The width of the eccentric part of the disk is about 0.2" or 
$0.7$pc so we would estimate that if the disk formed
from a disrupted cluster that its core radius
would have been smaller.  The total mass in
the most eccentric part of the disk is about
$10^6 M_\odot$ \citep{bacon} so we would require a progenitor
cluster of $r_0\sim 0.7$pc  and $\rho_0 \sim $ 
a few times $10^6 M_\odot{\rm  pc}^{-3}$.
For $\rho_0 \sim 4 \times 10^6 M_\odot{\rm  pc}^{-3}$
the disruption radius $r_d$ (estimated by equation \ref{inside})
is consistent with the location of the eccentric disk itself.

To provide a plausible progenitor, a cluster
must not only have a dense and compact core but
be massive enough to account for the mass of
the entire disk.
The extended, nearly circular part of the M31 stellar disk 
dominates its mass and has been estimated to be $3 \times 10^7 M_\odot$ 
\citep{peng,sambhus,bacon}.
This is massive compared to Galactic globular clusters,
presenting a problem for 
the globular cluster disruption scenario proposed 
by \citet{bekki}.
However extremely massive extra-galactic clusters have been identified.
For comparison, 
we show in Figure 1 the location of the massive
globular cluster (or perhaps core of
a dwarf galaxy) which is 40kpc from the nucleus of M31.
This cluster has 
a mass of $7-17 \times 10^6 M_\odot$ \citep{meylan},
a core radius $r_0 \approx 0.14"=0.52$pc,
and a core density $\rho_0 \approx 4.7\times 10^5 M_\odot {\rm pc}^{-3}$
\citep{meylan}.
G1 is massive
enough that it could have accounted for the mass
in the extended (and almost circular) part of M31's eccentric disk, though
its tidal radius is large, $\sim 200$pc, which implies
that much of the cluster would have disrupted well outside the location
of the eccentric disk in M31.  

The estimated progenitor core properties required
to form the M31 eccentric disk are surprising if
we compare them to Milky Way globular clusters and
nearly matched by the exotic and 
extremely massive extra-galactic cluster G1.
However, they are similar to what has been 
measured in galaxy nearby cores, such as M33, M32 and M31 itself, 
which we have also placed on Figure 1.
We see from this figure that
galaxy cores can have the high densities and compactness
to be eccentric disk progenitor candidates.
Moreover, they are usually more massive than globular clusters
and so can more easily account for the mass in M31's eccentric disk.

\begin{figure*}  
\vspace{10.0cm}
\includegraphics{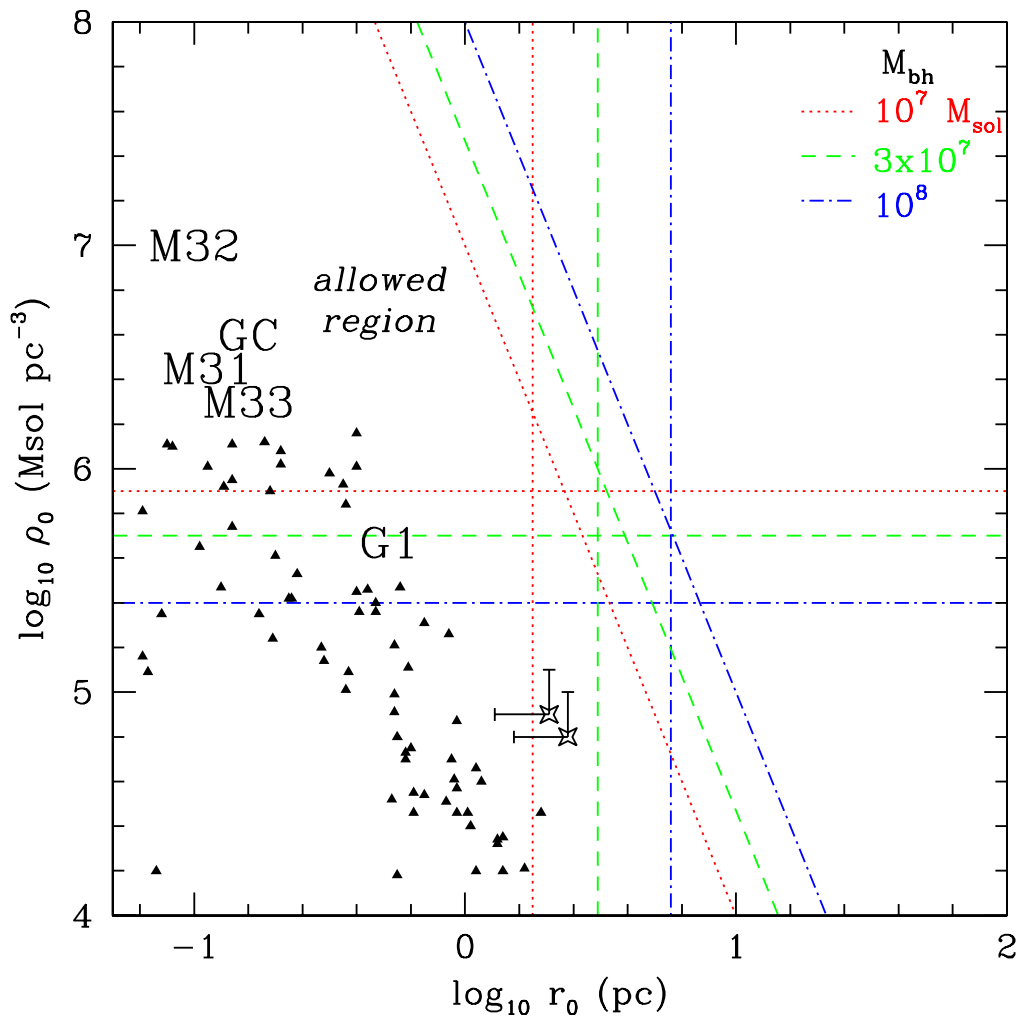}
\caption[]{
In this figure we outline the range of progenitor core radii and central
densities that would allow the creation of an eccentric disk near
a massive black hole.
Dotted lines are for a black hole mass of $M_{bh} = 10^7 M_\odot$.
The dashed lines are for $M_{bh} = 3\times 10^7 M_\odot$ and
the dot dashed lines for $M_{bh} = 10^8 M_\odot$.
Horizontal lines are set from the constraint $r_d < r_t$ where
we require that the disruption radius be within
the region where the black hole dominates the gravitational potential
(the sphere of influence).  
The allowed region for the progenitor core is that above the horizontal lines.
Vertical lines are set from the constraint  $r_0 < r_t$, requiring
that the core radius of the cluster is smaller than 
the region in which the black hole dominates the gravitational potential.
The allowed region is that to the left of the vertical lines.
The diagonal set of lines are set from the
constraint $M_d < M_{bh}$ where we require that the mass of the active
part of the disk $M_d\sim \rho_0 r_0^3$  be less than the black hole
mass.  
Parameter space for progenitor clusters that could result in 
the formation of an eccentric is limited to the upper leftmost area on the plot.
Galactic globular clusters \citep{harris} 
(assuming $M/L_V = 5$, solar units) are shown as triangles.
The G1 cluster in M31 (labeled as G1) is placed using parameters 
by \citet{meylan}.   
M31, M32 and M33 are placed on this plot with values derived by \citet{lauer98}.
In the case of M32 the core radius is only an upper limit.
The Galactic center (denoted GC) is placed on this plot based on 
values summarized by \citet{alexander}.
Limits on the core of two additional
power law galaxies, NGC 3377 and NGC 3115, with unresolved dense cores,
are shown as diamonds with bars extending outward to show
that these points are limits.  The values
for these two galaxies are based on values derived by \citet{faber}.
}
\end{figure*}

\subsection{NGC 4486's eccentric disk and its progenitor }

The other well known example with a double nucleus is
NGC 4486B, with 
a separation of 0.15" or 12 pc between the two isophotal peaks  
\citet{lauer96} and a black hole mass of
$M_{bh} \sim 10^9 M_\odot$ \citep{magorrian}.
The transition radius in this galaxy would be $r_t \sim 17$pc;
so again we find that the eccentric stellar disk lies within $r_t$.
From equation(\ref{inside}) and using 12 pc for
the disruption radius we estimate that the progenitor core
density would be  $\rho_0 \sim 6 \times 10^5 M_\odot {\rm pc}^{-3}$.
The width of the peaks comprising the disk (and hence the progenitor
core radius) is estimated to be 2--4 pc.
However, NGC 4486A is much more distant than M31,
and consequently the eccentric disk mass is probably
much higher than that of M31, probably exceeding $10^8 M_\odot$
and this is too massive for a globular cluster. 
However as in the case of M31's estimated progenitor, 
the central densities, core radius
and total mass are reasonable for a galaxy core.

\subsection{Comparison to the properties of galaxy cores}

In equation(\ref{rt_def}) we have assumed an isothermal
profile for the stellar density profile of the background galaxy.  
However, as shown by multiple studies (e.g., \citealt{faber,lauer98})
galaxy cores at small radii are seldom fit 
with a surface brightness profile proportional to $R^{-1}$
(where $R$ is the observed angular radius from the nucleus on
the sky) corresponding to an isothermal density profile
$\propto r^{-2}$.
Images from the HST have been
part of a major effort to classify the nuclear stellar profiles in 
early-type galaxies, resulting in the classification of light
profiles into two categories, galaxies with shallow inner
cusps, denoted `core-type' profiles and galaxies with 
`power-law' light profiles 
\citep{lauer95,faber}.    Power-law type galaxy cores tend
to have the steepest nuclear surface brightness profiles 
$\mu(R) \propto R^{-\gamma}$ with $\gamma$ nearly equal to 1 in
only the the most extreme cases.
Most of the galaxies studied by \citet{faber} had shallower profiles.
This implies that equation(\ref{r_t}) 
somewhat underestimates the transition radius in most galaxies, particularly
in the galaxies classified as core-type in which $\gamma$ is close to zero.
We can regard our estimated radius (equation \ref{r_t}) as a lower limit on 
the transition radius for all but the galaxies with the steepest
nuclear profiles.

We now consider the problem of two merging galaxies, both
with more complex stellar density profiles and both
with massive black holes.  We can approximately describe the primary as
having a transition radius given by equation(\ref{r_t})
(which is actually a lower limit as mentioned above).
If the secondary has a power-law form for its density
profile then it will not completely disrupt at a particular radius, 
like the non-singular isothermal sphere or King models.

One component of the
Nuker surface brightness profile \citep{faber} 
is the break radius, $r_b$,
where the slope of the surface brightness profile changes.
For profiles denoted core-type, within $r_b$, 
the surface brightness (and so density) rises much less
steeply with decreasing radius than outside it.  
Because the surface brightness for core-type galaxies
only increases slightly within $r_b$, we can 
associate the break radius and density estimated at this radius,
with the core radius $r_0$ and density  $\rho_0$ which we have used
in the previous sections.   This lets us determine
if the bulge or stellar component of the secondary galaxy will survive 
tidal truncation to within the transition radius of the primary.
A secondary with break radius exceeding
its own transition radius $r_b > r_t$ 
(likely to be true for luminous galaxies) will almost completely
disrupt when the tidal truncation radius reaches its break radius.
The black hole of the secondary will then spiral in to the nucleus
of the primary, leaving the stellar component of the secondary behind.

However for bulges and fainter ellipticals, 
we expect $r_b < r_t$.  For example,
M32's profile is steeply rising well within its
transition radius $r_t \sim 0.5$pc 
and increases in density to its last measured point \citep{lauer98}.
In this case the core would be continuously disrupted,
though we expect  
a change in the surface brightness profile of the resulting
disk set by the location at which 
the tidal truncation radius of the secondary
was equal to its own transition radius.
The continuous
stripping of the secondary in the presence of the
secondary's black hole 
may prevent the formation of a lopsided disk.

\subsection{Tidal disruption of a power-law galaxy}

The radius of tidal disruption can be estimated by
comparing the mean density of the object to that of
the object disrupting it
(e.g., \citealt{sridhar}).
At a distance $r$ from the nucleus of the primary galaxy, 
the mean density resulting from the black hole
is $\bar \rho(r) = M_{bh,1}/({4 \pi r^3 \over 3})$
where $M_{bh,1}$ is the mass of the primary's black hole.
We have assumed that $r<r_{t,1}$ where $r_{t,1}$ is
the transition radius of the primary, 
so the contribution from the primary's bulge is negligible.

For a secondary galaxy with a power-law law density profile 
(see \citealt{B+T} section 2.1e)
\begin{equation}
\rho(s) = \rho_a \left({a \over s}\right)^\alpha
\end{equation}
the mass integrated out to radius $s$ is  
\begin{equation}
M(s) = {4 \pi \rho_a a^\alpha \over (3-\alpha)} r^{3 -\alpha}
\end{equation}
so the mean density within this radius
\begin{equation}
\bar \rho (s) = 
    {3  \rho_a  \over (3-\alpha)} \left({ a \over s}\right)^{\alpha}.
\end{equation}
for radii $s> r_{t,2}$  outside the transition radius of the secondary.
We set the mean density of the primary within $r$ (due to the primary's
massive black hole) 
equal to the that of secondary (within stripping radius $s$),
$\bar\rho(s)=\bar\rho(r)$,
and find that 
\begin{equation}
\label{r3}
r^3 \approx { M_{bh} (3 -\alpha) \over 4 \pi \rho_a}
   \left({s \over a}\right)^\alpha.
\end{equation}
When the secondary is at a distance $r$ from the nucleus
of the primary, it will be stripped out to a distance
$s$ from its own nucleus, where $s$ is related to $r$ by
the previous equation.

The stripping radius, $s$, is the same size as the distance
from the primary's nucleus, $r$, at a radius
\begin{equation}
\label{def_re}
r_e = \left[
     { M_{bh,1} (3 -\alpha) \over 4 \pi \rho_a a^3}
    \right]^{1 \over 3-\alpha} a
\end{equation}
where $M_{bh,1}$ is the mass of the primary's black hole.
When $s$ estimated from equation(\ref{r3}) exceeds $r$
then the secondary engulfs the nucleus of the primary.

For a disrupted secondary galaxy core to produce an eccentric
disk, we expect that the radius at which the secondary
engulfs the nucleus of the primary, $r_e$, should
be smaller than the  transition radius of the primary ($r_{t,1}$).
In this case, the core of the secondary will survive intact
within the transition of the secondary.
A reasonable condition for formation of
an eccentric disk via the disruption of a galaxy with a power-law
density profile should be 
\begin{equation}
r_e < r_{t,1}.
\end{equation}
Using equations(\ref{def_re},\ref{r_t}) this is equivalent to 
\begin{equation}
\rho_a > M_{bh,1}^{\alpha - 1 \over 2} a^{-\alpha} \kappa^{\alpha -3}
{(3 -\alpha) \over 4 \pi}
\end{equation}
where $\kappa = 5.7 \times 10^{-4} {\rm pc} M_\odot^{-1/2}$.
For an isothermal profile, 
the limit on the density at $a=1$pc
\begin{equation}
\label{rho_a_lim}
\rho_{a=1{\rm pc}} > 5 \times 10^6 M_\odot {\rm pc}^{-3} 
   \left({M_{bh,1} \over 10^7 M_\odot}\right)^{1/2}   ~~~~~ {\rm for} ~\alpha=2.
\end{equation}

\subsection{Existing simulations of galaxy mergers}

With the exception of \citet{merritt,holley},
few simulations have been carried out with realistic stellar density
profiles and central black holes.
\citet{merritt} presented N-body simulations 
of two merging galaxies, the primary with a shallow
central density profile and the secondary with a steeper profile
at the nucleus.  They did simulations with and without
massive nuclear black holes to illustrate the difference in
remnant profiles following the merger.
The density profiles adopted for their simulations
were generated from the family introduced by \citet{dehnen93}.

To explore the role of the black holes on the disruption
of the secondary in their simulations 
we estimate the transition radii for each galaxy.
At small radii, $r<a$, both galaxies were power-law in form 
\begin{equation}
\label{rho}
 \rho(r) = { (3 - \alpha) M   \over 4 \pi a^3 }
\left({a \over r}\right)^\alpha
\end{equation}
where $M$ was the mass of the galaxy and $a$ was its scale length.
\citet{merritt} used $\alpha_1 = 1$ for the primary, and $\alpha_2=2$
for the secondary (subscripts here refer to the primary
and secondary galaxy respectively).  
We estimate the transition radius for each galaxy
(sphere of influence) by calculating 
the radius at which the integrated stellar mass within 
equals that of the central black hole.
For the primary galaxy 
\begin{equation}
r_{t,1} \approx a_1 \left({M_{bh,1} \over M_1}\right)^{1/2}
= 0.035 a_1
\end{equation}
where we have integrated equation(\ref{rho}) and 
used the ratio $M_{bh}/M = 0.0012$ (that adopted by \citealt{merritt}).
For the secondary galaxy we estimate
\begin{equation}
r_{t,2} \approx a_2 \left({M_{bh,1} \over M_1}\right)
= 0.0012 a_2.
\end{equation}
Because the transition radius of the secondary was so small,
its black hole should not have affected the simulations (they
were probably not well resolved on that scale).

Both simulations produced similar remnants at radii larger
than 0.1 times the primary's scale length.  The change
in the surface brightness profiles at this radius was
probably due to the break radius (scale length at which the 
density profile changes slope) of the secondary.

For the simulations containing both massive black holes,
a change in radial density profile is seen in the remnant 
at approximately the primary's transition radius
(see their figure 2d.)
Using their profiles, we estimate that $r_e \sim 0.01$ times
the scale length of the secondary, which is consistent
with the survival of a fraction of 
the secondary's core to well 
within the transition radius of the primary.
The constraints developed in the previous section
were not violated ($r_e < r_{t,1}$), so this type of simulation 
might have been capable of forming an eccentric disk
in the remnant.
If the simulation did not form one, then simultaneous
disruption of the densest part of the cluster might
be required to form an eccentric disk.
This would occur when the core has a
well defined core radius (and nearly constant density within), 
and may not occur when the secondary has a 
steeply increasing density profile all the way to its nucleus.

In the simulations lacking the massive black holes (see their
figure 2h), the secondary did not disrupt within its 
break or scale radius.  This follows because
the mean density of the secondary significantly exceeds the
mean density of the primary within their scale radii, consequently 
the primary could not have disrupted the secondary within 
the secondary's scale radius.

The merger simulations of \citet{holley} were done with a quite massive
primary black hole $M_{bh} > 10^9 M_\odot$ and lacked a secondary
black hole.  From their initial density profiles we estimate that
the size of the secondary at its disruption radius
in their 10:1 simulation was only about a third the disruption
radius itself.   This implies that a disk could be formed
from the secondary, following disruption.  
\citet{holley} reported that
their simulations produced a spinning remnant disk, however it was 
was thick, consistent with our crude estimate of the secondary's
largish size during disruption.
Much of the secondary survived to within the transition radius
of the pimary's black hole (about 30 pc). 
It is possible that a denser secondary (better satisfying the conditions
in eqns 18--20) could result in a thinner stellar disk which
might then be more likely to be lopsided.
Alternatively, a secondary with a shallow core  
(that would would disrupt all at once, 
as simulated by \citealt{bekki})
might be required for eccentric disk formation.

\section{Discussion and Summary}

In this paper we have considered the possibility
that a single disruption event could result
in the formation of an eccentric stellar disk such
as are found in M31 and possibly NGC 4486B.
We explore the scenario of a galaxy core
which is stripped as it
spirals in toward a nuclear massive black hole 
until it reaches a critical radius, denoted
here as $r_d$, the disruption radius.
In the simplest model of an isothermal sphere,
the disruption radius 
is set by the cluster core or King radius and central
density.    We suggest that an eccentric disk
cannot form unless the core of the galaxy or stellar cluster
is disrupted within the sphere
of influence of the massive black hole, referred to
here as the transition radius, $r_t$, and the core radius
is smaller than the radius at which it disrupts.
To succeed in making an eccentric disk, we find
that the progenitor cluster core must be denser 
and more compact for less massive black holes.

For the eccentric disk in M31, the progenitor cluster core density 
must be greater
than $10^5 M_\odot {\rm pc}^{-3}$ and core radius
smaller than a pc.
From the radius, width and mass of the most eccentric region 
of the disk we estimate
that the density of the core was a few times 
$10^6 M_\odot {\rm pc}^{-3}$ and the core radius was
of order a half pc.  

Massive extragalactic globular clusters such as G1 might be dense,
compact and massive enough to be an eccentric disk progenitor,
however they are probably too diffuse to account for the large 
disk mass in M31 within a few parsecs of its black hole.

Dense bulges such as found in M31 itself, and in lower luminosity
ellipticals galaxies such as M32, are dense and compact enough
that they could be similar to 
progenitors for M31's and NGC 4486B's eccentric disks.
Nuclear star clusters, such as found in M33,  may also
be dense and compact enough to be progenitors.
Though the mass of M33's cluster (a few million $M_\odot$)
is too small to be a progenitor for M31's eccentric disk,
the nuclear stellar clusters observed by \citep{boker2002}
in late-type galaxies have half light radii of order 5pc,
range in their estimated masses from $10^6-10^8 M_\odot$,
and so could be progenitors for eccentric disks following
disruption by a massive black hole.
Because of the large mass of bulges and nuclear star
clusters at small radii, it would be easier to account for the high masses
of the two known eccentric disks by disrupting them, 
than possible by disrupting a globular cluster.
While the massive black hole in M32 might prevent a progenitor
similar to M32 from forming an eccentric disk, 
nuclear clusters such as found in M33 can lack massive black holes 
and so might provide better progenitor candidates.
To date the high angular resolution of HST has resolved extremely
high stellar densities or order $10^6 M_\odot{\rm pc}^{-3}$
in only the nearest galaxy bulges.
Until higher angular resolution observations are available,
we will not know if such high density galaxy cores are common.

The two galaxies with double nuclei, M31 and NGC 4486B,
exhibit only moderate color variations in their nuclei 
\citep{lauer96,lauer98}, a situation that 
could be a natural consequence of a scenario that involves
the merging of galaxy bulges (proposed here), and is more difficult to explain
with a scenario that forms a younger disk in situ
(an ingredient of the formation scenarios discussed 
by \citealt{bacon,touma,taga,jacobs}).

The scenario proposed here is based on a simple tidal
disruption argument and can be most quickly tested with N-body simulations 
such as have been carried out by \citet{merritt,bekki,holley}.
The simulations of
\citet{bekki} established that the disruption of a cluster
could result in the formation of an eccentric disk, and \citet{merritt,holley}
have carried out simulations based on realistic galaxy profiles
and with massive black holes.    \citet{holley} established
that a spinning nuclear stellar disk can be a remnant.
It remains to be seen whether 
the merger of two galaxy cores can result in
the creation of an eccentric stellar disk. 

We suspect that the merger of a primary galaxy containing a very massive
black hole with a secondary with a `core-type' or shallow
central surface
brightness profile, and significantly lower mass black hole,
would be most likely 
to form an eccentric disk with N-body simulations.
For the core-type galaxies, the break radius and density at this
radius provide an equivalent for the King or core radius 
and central density used in Figure 1 
and so can be used to estimate the likelihood
that the core can disrupt to form an eccentric disk.

If the merger of galaxy bulges can result in the formation
of an eccentric disk, then their formation
would be a natural consequence of hierarchical galaxy
formation, and can also be used to probe the properties
of the parents of galaxies which contain them.

\acknowledgments
This work was initiated by discussions with Joel Green and  Rob Gutermuth
in the class Astronomy 552 at the University of Rochester during the fall
of 2002. We thank Chien Peng, Ari Laor and Eric Emsellem for
helpful discussions and correspondence.

{}

\end{document}